\documentclass{aastex62}

\usepackage{multirow}

\begin{document}

\newcommand{\rbs}{{\small RBS\,315}}
\newcommand{\swift}{\small {\it Swift}}
\newcommand{\xrt}{{\small {\it Swift}/XRT}}
\newcommand{\xmm}{{\small \it XMM-Newton}}
\newcommand{\nustar}{{\small \it NuSTAR}}
\newcommand{\cstat}{$C_{\mathrm stat}$/d.o.f.}
\newcommand{\rchi}{$\chi ^2$/d.o.f.}

\title{X-Ray Spectrum of RBS 315: Absorption or Intrinsic Curvature}

\author{Sivan Ben Haim}
\affiliation{Department of Physics, Technion, Haifa 32000, Israel}
\email{sivan.bhaim@gmail.com}

\author{Ehud Behar}
\affiliation{Department of Physics, Technion, Haifa 32000, Israel}


\author{Richard F. Mushotzky}
\affiliation{Department of Astronomy, University of Maryland College Park}
 
\begin{abstract}

X-ray absorption of high-redshift quasars is enigmatic, because it remains
unclear where in the universe the absorbing gas is. If absorption
occurs near the high-z host, it could help us understand early stages of galaxy formation.
If it is in the intergalactic medium (IGM), it provides a unique way to probe this elusive baryon component.
We report on observations
of one of the brightest X-ray sources at a high redshift, RBS 315
($z=2.69$). 
Despite several previous analyses, no definite conclusion as to the source of the
curvature in its spectrum, whether absorption or intrinsic, could be reached.
We present observations by \xmm\ (EPIC and RGS) as well as  \nustar\ and \xrt . 
The \xmm\ spectra of this source are of unprecedented quality.
A purely statistical analysis of the CCD spectra yields no clear results - the spectrum is as likely to be
photo-electrically absorbed as it is to be curved at the source, and no constraint on the position of the absorber can be obtained.
Assuming absorption governs the spectral curvature, the lack of absorption lines in the grating spectra indicates the absorber is not well localized in redshift space, and could be dispersed over the cosmological scales of the IGM.
Intrinsic curvature, however, can not be unambiguously ruled out.
\end{abstract}


\section{Introduction} \label{sec:intro}

X-ray absorption of high-redshift ($z>2$) sources holds the key to
understanding cosmological feedback when accretion activity was most
intense.
It facilitates the search for baryons and especially metals in the intergalactic medium (IGM). 
Although current CCD X-ray spectra do not reveal the redshift of the absorption, 
the gradual increase of X-ray opacity with $z$ towards high-$z$ sources suggests a possible diffuse-IGM origin  and evidence for the missing baryons of the universe \citep{key-8, Campana12, Campana15, Starling13}.
A comprehensive survey of high-$z$ AGNs revealed mixed results.
Despite the mean optical depth increasing with redshift, many AGNs are not sufficiently absorbed \citep{key-4}.
A clumpy and partially ionized IGM, which is expected from simulations, could explain such results \citep[e.g.,][]{Starling13, Campana15}.  
Conversely, the curvature of AGN spectra, which is attributed to foreground absorption could also be due to intrinsic curvature.
This was deemed unlikely, because most of the sources show a break around the observed energy of 1\,keV, regardless of quasar redshift \citep{key-4}.
Direct detection of IGM X-ray absorption lines is paramount to conclusively claiming detection of these missing baryons. 
This task has proven to be extremely difficult \citep[e.g.,][]{Nicastro05, Nicastro18}.

Among the high-$z$ sources, the FSRQ \rbs\ ($z=2.69$) is one of the brightest, with an X-ray flux of $F_{2-10\ \mathrm{keV}}=1.6(1.08)\times{}10^{-11}\  \mathrm{erg\ s^{-1}\ cm ^{-2}}$ as observed with \xmm\ ({\it Suzaku}) \citep{key-1}, and one of the most absorbed with an optical depth $\tau$(at $0.5$ keV) $=1.08\pm0.06$ \citep{key-8}.
RBS 315 is almost twice brighter than the most luminous quasars in the local universe; For comparison, the X-ray flux from
PDS 456 ($z=0.184$) in its high state is $F_{2-10\ \mathrm{keV}}=0.6\times{}10^{-11}\ \mathrm{erg\ s^{-1}\ cm ^{-2}}$.
For this reason, \rbs\ is an excellent candidate for a deep grating observation that could potentially reveal absorption features, and thus shed new light on the nature of the absorber.
Blazars in particular are good targets for IGM searches, as their powerful jet would most likely remove any host material along the line of sight. 
Indeed, recently \citep{Arcodia18} analyzed a sample of $z > 2$ flat radio spectrum blazars, reaching the conclusion that extragalactic absorption is the preferred explanation for the soft X-ray spectral turnover.

\citet{key-2} analyzed the 20\,ks \xmm\ spectrum of \rbs\ obtained in July 2003
and measured an extragalactic neutral hydrogen column density of $N_{\mathrm H(z)}=(1.62\pm 0.09) \times10^{22}\ \mathrm{cm^{-2}}$ if the absorber is at $z = 2.69$ and is neutral.
The column is slightly higher if the absorber is partially ionized.
Similarly, \citet{key-1} found $N_{H}= (1.65\pm 0.09) \times10^{22}\ \mathrm{cm^{-2}}$ for the same observation.
These measurements were done while assuming older metal abundances in the absorber.
The assumed abundances make a big difference in the deduced equivalent $N_{\mathrm H}$.
In this case, using the present abundances \citep{Wilms00} results in $N_{\mathrm H(z)} = (2.9 \pm 0.2) \times10^{22}\ \mathrm{cm^{-2}}$.
However, based on a {\it Suzaku} observation in July 2006, \citet{key-2} measured $N_{\mathrm H(z)} = (2.7\pm{}0.4) \times10^{22}\ \mathrm{cm^{-2}}$.
A possible absorption variability argues against an IGM origin \citep{key-1}, but as we show below variable spectral slopes and high-energy breaks could bias the measured column density.

Despite those efforts, reliable diagnostics of absorber redshift are not possible with CCD spectra, and thus all of the different scenarios remained viable. \citet{key-1} suggested
that the spectral curvature could be intrinsic and not due to absorption.
\citet{key-9} observed \rbs\ with \xrt\ and \nustar ,
and they too concluded that the curvature is more likely intrinsic
to the source, rather than resulting from an absorber.

In this work, we aim to exploit the deepest observation of \rbs\ to date, a 166\,ks \xmm\ exposure we obtained in 2013, to test whether the high-statistics CCD (EPIC) spectra can distinguish between absorption and intrinsic curvature, and to see if the high resolution  ($\Delta \lambda \approx $70 m\AA ) grating (RGS) spectrum reveals discrete absorption features.
We also use the aforementioned \nustar\ and \xrt\ observations
 in order to reach a more definitive conclusion as to the source of the strong curvature in the spectrum.

\section{Data Analysis} 
\label{sec:method}

All of the observations used in this paper are listed in Table 1.
The \xmm\ data were obtained from the HEASRAC Archive\footnote{https://heasarc.gsfc.nasa.gov/docs/archive.html}.
For the EPIC PN detector, we used the $\rm{epn\_ff20\_sdY9.rmf}$ response matrix. 
For the EPIC MOS1 and MOS2 detectors, we used $\rm{m1\_e13\_im\_pall\_c.rmf}$ and $\rm{m2\_e13\_im\_pall\_c.rmf}$ respectively. 
The \xrt\ data were processed using CALDB\footnote{https://heasarc.gsfc.nasa.gov/docs/heasarc/caldb/install.html}
and HEASOFT's XSELECT tool \footnote{https://heasarc.nasa.gov/docs/software/lheasoft/ftools/xselect/node1.html}.
Background was estimated from an annular area of a radius of $\sim50$ pixels
around the source and then subtracted from the central source region using ds9\footnote{http://ds9.si.edu/site/Home.html}. 
The statistical quality of these spectra is very far from that of EPIC, hence we grouped the data, using the grppha routine to have at least 8 counts in each bin. This allowed us to use $\chi ^2$ and compare with the $C_{\mathrm stat}$ fitting on the un-binned spectra.

\begin{table}
\begin{centering}
\begin{tabular}{|c|c|c|c|c|}
\hline 
\multirow{2}{*}{Instrument} & Energy Range  & \multirow{2}{*}{Date} & Duration & Total Counts\tabularnewline
 & {[}keV{]} &  & {[}ksec{]} & {[}$10^{4}${]}\tabularnewline
\hline 
\hline 
\multirow{2}{*}{\xmm /EPIC} & \multirow{2}{*}{$0.3-10$} & Jan 2013 & $86$ & $32.4$\tabularnewline
\cline{3-5} 
 &  & Jan 2013 & $80$ & $37.4$\tabularnewline
\hline 
\multirow{2}{*}{\xmm /RGS} & \multirow{2}{*}{$0.3-2$} & Jan 2013 & $101$ & $1.5$\tabularnewline
\cline{3-5} 
 &  & Jan 2013 & $96$ & $1.3$\tabularnewline
\hline 
\multirow{2}{*}{\xrt } & \multirow{2}{*}{$0.3-10$} & Dec 2014 & $5$ & $0.1$\tabularnewline
\cline{3-5} 
 &  & Jan 2015 & $5$ & $0.08$\tabularnewline
\hline 
\multirow{2}{*}{\nustar} & \multirow{2}{*}{$3-70$} & Dec 2014 & $32$ & $2.0$\tabularnewline
\cline{3-5} 
 &  & Jan 2015 & $37$ & $1.6$\tabularnewline
\hline 
\end{tabular}
\par\end{centering}
\caption{Observations Log}

\end{table}


We used XSPEC\footnote{https://heasarc.gsfc.nasa.gov/xanadu/xspec/} to fit different candidate models.
The two types of models, which fit the X-ray spectra of \rbs\ well are an absorbed power law, and a broken power law.
The {\it tbnew} model was used to represent the Galactic absorption, known to have 
$N_{H}=9.26\times10^{20}\ \mathrm{cm^{-2}}$ \citep{key-3} in the direction of \rbs . 
For the excess absorption we used an additional absorber model {\it tbnew} 
(\citet{Wilms00} and \footnote{http://pulsar.sternwarte.uni-erlangen.de/wilms/research/tbabs/}) and its solar abundance values.
An intrinsic curvature model was also tested by fitting a broken power law.

\section{Results} \label{sec:results}

We started with the highest-quality (by far) spectrum available of \rbs , namely the 2013 \xmm /EPIC observation.
We fitted the spectra of EPIC PN, MOS1, and MOS2, simultaneously between 0.3~- 10~keV.

\subsection{\xmm /EPIC} \label{subsec:epic}

We find that an absorbed power law model fits the spectrum reasonably well.
The spectral index is about $\Gamma = 1.4$, and changes slightly with the redshift of the extragalactic absorber.
The fit is slightly better (\rchi = 1.11, \cstat = 1.12) for a $z=0$ absorber than for one at the host (\rchi = 1.20, \cstat = 1.25), but the null hypothesis probability in both cases is unacceptable (4E-8 and 1E-21, respectively).
This is due to the known degeneracy between the redshift and the column density $N_\mathrm{H}$, which is demonstrated in Figure\,1.
Since the photo-electric cross section decreases with energy, a higher-redshift absorber requires higher column density to produce the same attenuation (optical depth) effect.
Indeed, the best-fit absorption models yield a column density at the host of ($3.19\pm 0.05) \times10^{22}\ \mathrm{cm^{-2}}$, which is marginally consistent with $(2.9 \pm 0.2) \times10^{22}\ \mathrm{cm^{-2}}$ obtained from the 2003 EPIC spectrum \citep{key-4}, while $N_{\rm H} = (1.6 \pm 0.02) \times10^{21}\ \mathrm{cm^{-2}}$ is sufficient to fit a $z=0$ absorber.
As a result of the degeneracy, the absorber, even if it exists and is well localized in redshift space, can not be pinpointed to a specific redshift, based on statistics alone (Fig.\,1).
We note that allowing somewhat higher Galactic absorption, per the molecular content \citep{Willingale13}, could reduce the redshifted column density.

\begin{figure}[h]
\label{NHz}
\includegraphics[width=0.7\columnwidth]{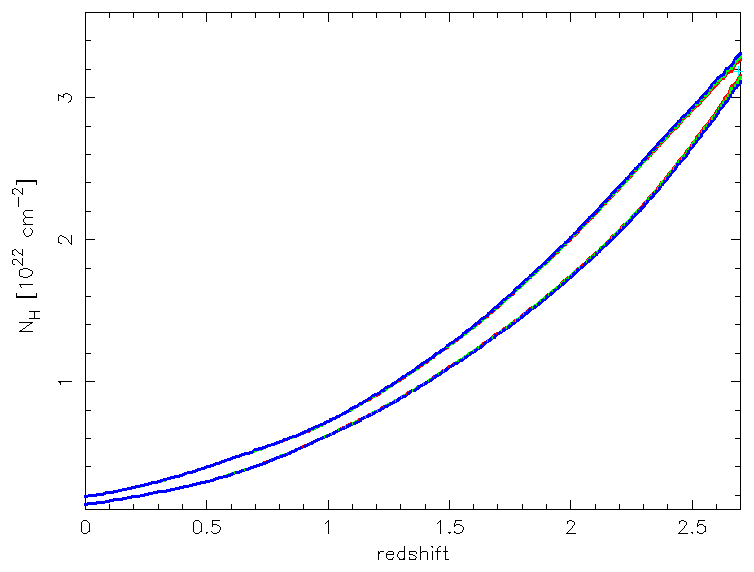}
\caption{Confidence region of column density vs. redshift based on the absorbed power law fitted to the 2013 EPIC spectrum. The three contours ($\Delta \chi ^2 = 2.3, 4.61, 9.21$ corresponding to 1, 2, 3 $\sigma$ confidence levels) are barely distinguishable due to the steep dependence of $\chi ^2$ on $N_{\mathrm H}$. The degeneracy between the two parameters is a result of the decreasing photo-absorption cross section with energy (absorber redshift).
As a result, it is impossible to determine the redshift of the absorber from the fit.
}
\end{figure}

We also attempted to fit a broken power-law model to the EPIC spectrum, and reached \rchi = 1.28 or \cstat =1.26, which is   statistically inferior to the absorption model above.
The best-fit break energy is $E_{b}=1.13\pm0.01$\,keV, which is somewhat higher than $0.95 \pm 0.05$\,keV found from the 2003 EPIC spectrum \citep{key-4}.
The two photon indices are $\Gamma_{1}=0.27 \pm 0.02$ and $\Gamma_{2}=1.37\pm0.005$. 
The high-energy power law is similar to that found above for an absorption model, while the flat low-energy slope mimics the effect of absorption.
All best-fit parameters are shown in Table\,2.

\begin{table}[h]
\begin{centering}

\begin{tabular}{|c|c|c|c|c|c|c|c|}
\hline 
\multirow{2}{*}{Model} &
\multirow{2}{*}{Redshift} & $N_\mathrm{H}$ & \multirow{2}{*}{$\Gamma_{1}$} & \multirow{2}{*}{$\Gamma_{2}$} & $E_{b}$ &
\multirow{2}{*}{\rchi} & \multirow{2}{*}{Degrees of Freedom}\tabularnewline
&  & {[}$10^{22}\: \mathrm{cm}^{-2}${]} &  &  & {[}keV{]} &  & \tabularnewline
\hline 
\hline 
Absorber &
$0$    & $0.161\pm0.002$    & $1.452\pm0.005$ & \nodata    & \nodata &
$1.11$ & $5197$
\tabularnewline \hline
Absorber &
$2.69$ & $3.19\pm0.05$      & $1.398\pm0.004$ & \nodata    & \nodata &
$1.20$ & $5197$
\tabularnewline \hline
Broken Power-Law &
\nodata & \nodata & $0.27\pm0.02$   & $1.369\pm0.005$ & $1.13\pm0.01$ &
$1.28$ & $5196$
\tabularnewline \hline
\end{tabular}

\vspace{0.5cm}

\caption{Best-fit parameters for models fitted to the 2013 \xmm\ EPIC spectrum.}

\par\end{centering}
\end{table}

\subsection{\xrt\  and \nustar\ } \label{subsec:swift_nustar}

\xrt\ and \nustar\ spectra of \rbs\ were taken together during two epochs: in 2014 Dec and in 2015 Jan.
Therefore, we fitted models to both instruments at each epoch, in the full 0.3--70 keV band.
We used both $C_{\mathrm stat}$ fitting, and $\chi ^ 2$ fitting on binned spectra (see Sec.\,\ref{sec:method}).
Due to the moderate statistical quality of the data all models yield \rchi\ $\approx 0.9$ and \cstat $\approx0.8 $, indicating that uncertainties are likely still not totally gaussian, despite the binning.
Similar to \citet{key-9}, we find that apart from the curvature around 1\,keV, the spectrum shows a break at high energy.
This finding is made possible owing to \nustar\ that constrains the power law spectra well above 10\,keV.
\citet{key-9} found the break to be at $\sim 4.7$\,keV, while we find it at slightly higher energies.
However, since both breaks result in \rchi\ values below unity, we do not claim to determine the break energy with high confidence, only to point out that it is much higher than where the putative absorption effect is dominant, around 1\,keV.
\citet{Arcodia18} also found that \rbs\ is well fitted with both spectral curvature and extragalactic absorption, though they prefer a log parabola continuum. The presence of the break is demonstrated in Figure\,2, where we plot the best fitted broken power law model on the Dec \nustar\ spectrum, after fixing the high-energy slope to match the low-energy slope.

We conclude therefore that the high-energy break in the power law is independent of the absorption.
Once the high-energy break is included in the \xrt\ + \nustar\ fitted models, we can also include an absorber, which further reduces \rchi\ (or \cstat\ for the un-binned data) by a few percent. The effect of the absorber is shown in Figure\,2, where the extragalactic column density in the best-fit model for the Dec \nustar\ data is set to zero.
The best fit parameters are listed in Table\,3, for both models in which the break energies were fixed to those of \citet{key-9}, and models in which it was fitted for.

\begin{figure}[h]
\includegraphics[width=0.5\columnwidth]{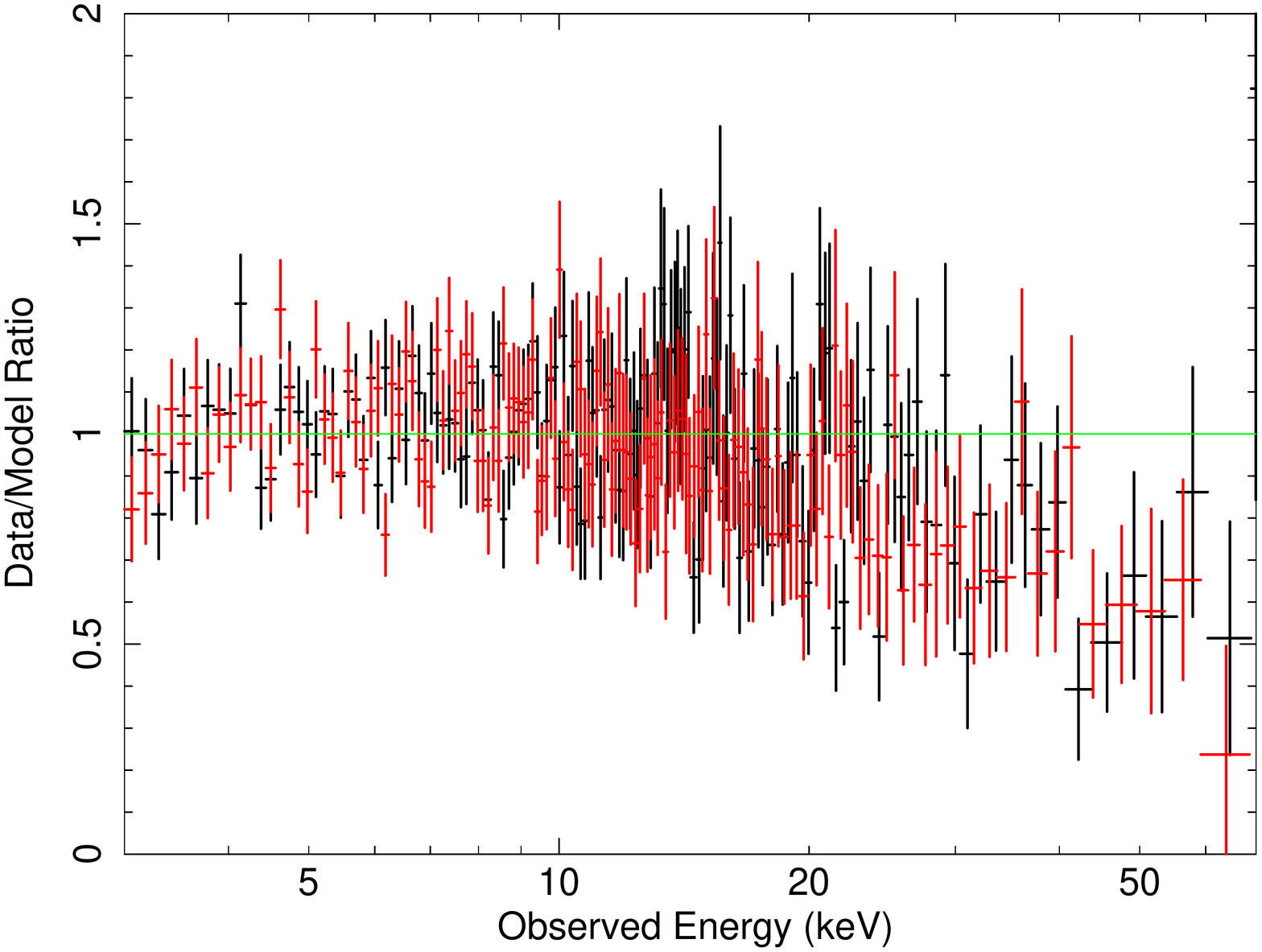}
\includegraphics[width=0.5\columnwidth]{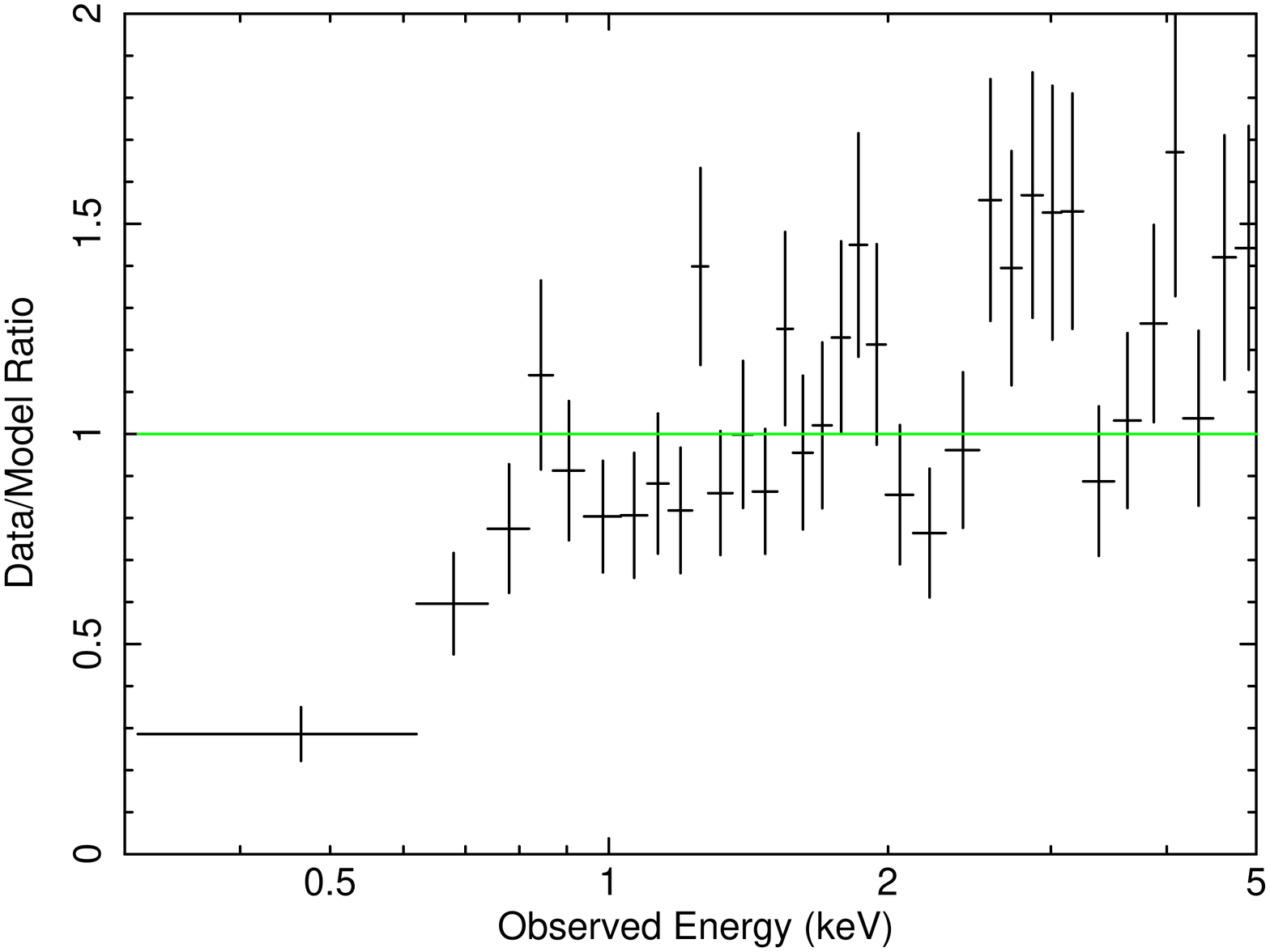}
\caption{The ratio of the data to the best-fit model when the $\approx 10\mathrm {keV}$ break in the power law is removed (\textit{left} shown on the \nustar\ spectra), or extragalactic absorption is removed (\textit{right} shown on the \xrt\ spectrum). This demonstrates the significance of these two components.}
\end{figure}

\begin{table}[h]
\begin{centering}

\begin{tabular}{|c|c|c|c|c|c|c|}
\hline 
\multirow{2}{*}{Observation} &
$N_{H}$ & \multirow{2}{*}{$\Gamma_{1}$} & \multirow{2}{*}{$\Gamma_{2}$} & $E_{b}$ &
\multirow{2}{*}{\rchi{}} & \multirow{2}{*}{Degrees of Freedom}\tabularnewline
 & {[}$10^{22}\: \mathrm{cm}^{-2}${]} &  &  & {[}keV{]} & & \tabularnewline
\hline 
\hline 
NuSTAR and XRT (Dec 2014) &
$3.2_{-1.3}^{+1.6}$ & $1.2\pm0.1$ & $1.59\pm0.03$ & $4.77$ (fixed) &
$0.89$ & $1182$
\tabularnewline \hline
NuSTAR and XRT (Dec 2014) &
$4.4_{-1.3}^{+1.5}$ & $1.41_{-0.12}^{+0.06}$ & $1.73_{-0.10}^{+0.15}$ & $9.8_{-3.1}^{+4.4}$ &
$0.87$ & $1181$
\tabularnewline \hline 
NuSTAR and XRT (Jan 2015) &
$4.3_{-2.4}^{+3.2}$ & $1.6\pm0.1$ & $1.70\pm0.03$ & $4.69$ (fixed) &
$0.96$ & $1129$
\tabularnewline \hline 
NuSTAR and XRT (Jan 2015) &
$4.0_{-1.8}^{+2.3}$ & $1.60_{-0.06}^{+0.05}$ & $1.86_{-0.10}^{+0.17}$ & $11.2_{-2.7}^{+3.9}$ &
$0.95$ & $1128$
\tabularnewline \hline 
\end{tabular}
\caption{Broken power law models with absorption (at the host) fitted to the \xrt\ + \nustar\ spectra. We fixed the break energies to those found by \citet{key-9} (around 4.7\,keV), and also fitted them again here. All models are over-constrained as indicated by the \rchi\ values, which are all below unity. 
}

\end{centering}
\end{table}

We find that including a broken power law say at 4.7\,keV in the models for the EPIC spectra (see Sec.\,\ref{subsec:epic}) slightly improves its fit as well. 
The absorber retains its high column density (assuming at $z = 2.69$) of now ($3.02\pm 0.05) \times10^{22}\ \mathrm{cm^{-2}}$ (compared to $3.19\pm 0.05 \times10^{22}\ \mathrm{cm^{-2}}$ without the break, see Table\,2), with a slightly reduced \rchi\ = 1.16 (c.f., 1.20 without the break). Regardless of the high-energy break in the power law, the extragalactic absorption is absolutely required to fit the EPIC spectra.

\section{Discussion}
\label{sec:discussion}

One way to distinguish between a local absorber at the host, and an inter-galactic one is variability.
An absorber that changes within a few years can not be on inter-galactic length scales, and must be ascribed to the host, or to intrinsic variability of the source (the no-absorption model).
We note that all column densities quoted in Tables 2 and 3, assuming an absorber at $z= 2.69$ are consistent within the errors.
Since column density is somewhat degenerate with spectral slope; higher column density (harder spectrum) could mimic a steeper slope, and vice versa. 
For this purpose we compare the three epochs of \rbs\ in Figure\,3, which shows the confidence regions of the absorber column density vs. the power law spectral index in the three data sets discussed above.
The power of a large telescope like \xmm\ is demonstrated in the tight EPIC contours in the figure.
Yet, the EPIC column density measurement is totally consistent with the (much larger) \xrt\ + \nustar\ uncertainties.
These contours show that if there is an absorber it did not change significantly between 2013 Jan (actually the same column of 2003) and 2015 Jan.
This is consistent with previous reports. While \citet{key-9} find that the source varies on the short time scales between \swift\ observations, \citet{Arcodia18} conclude that fixing the absorber, while the continuum varies, provides a good fit to the data, as we do.

\begin{figure}[h]
\includegraphics[width=0.8\columnwidth]{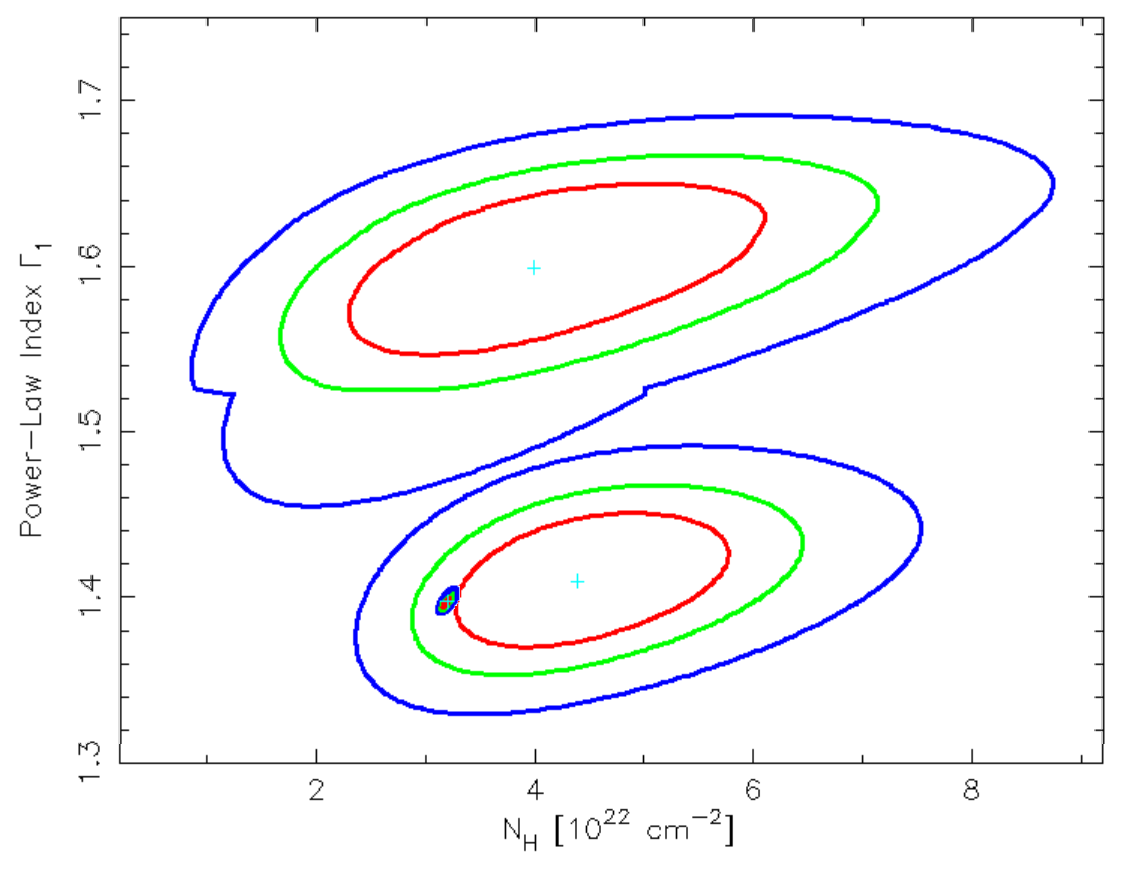}
\caption{Comparison of column density uncertainties from the EPIC and two \nustar + \xrt observations of \rbs . Note the minuscule EPIC contours around $N_{\mathrm H} = 3 \times10^{22}\ \mathrm{cm^{-2}}$ and $\Gamma = 1.4$. All three column density measurements are consistent with each other, despite the varying spectral index, leaving the intergalactic absorber as a viable possibility.}
\end{figure}

Furthermore, Figure\,4 shows the three data sets (EPIC and two \xrt\ + \nustar\ epochs) compared to a $\Gamma = 2$ power law. 
Despite the different slopes, the sense of absorption below $\sim 1$\,keV appears to be similar, giving us confidence that the absorber could have the same column density in all epochs.

\begin{figure}[h]
\includegraphics[width=6.5cm]{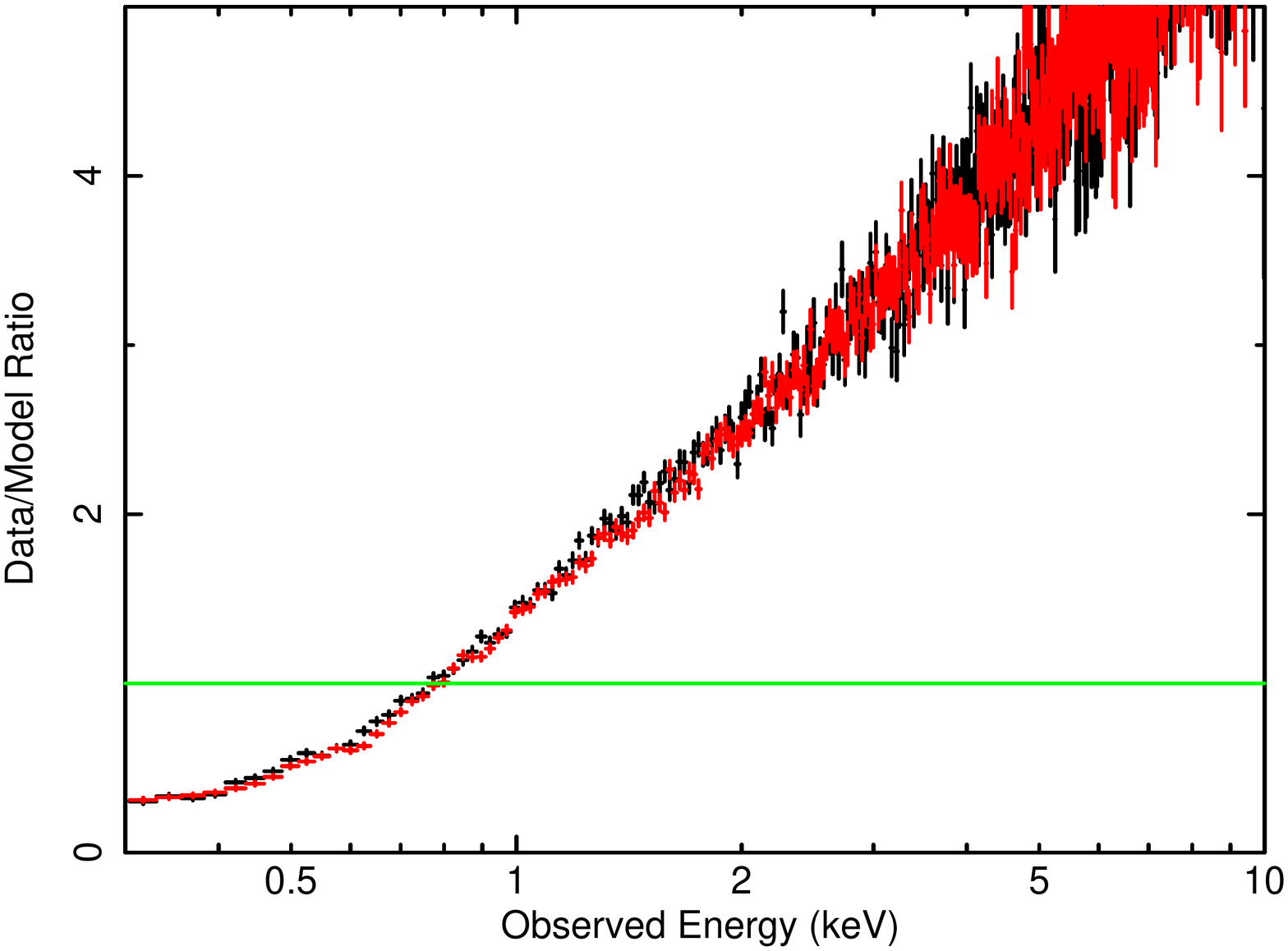}
\hspace{-0.8cm}
\includegraphics[width=6.5cm]{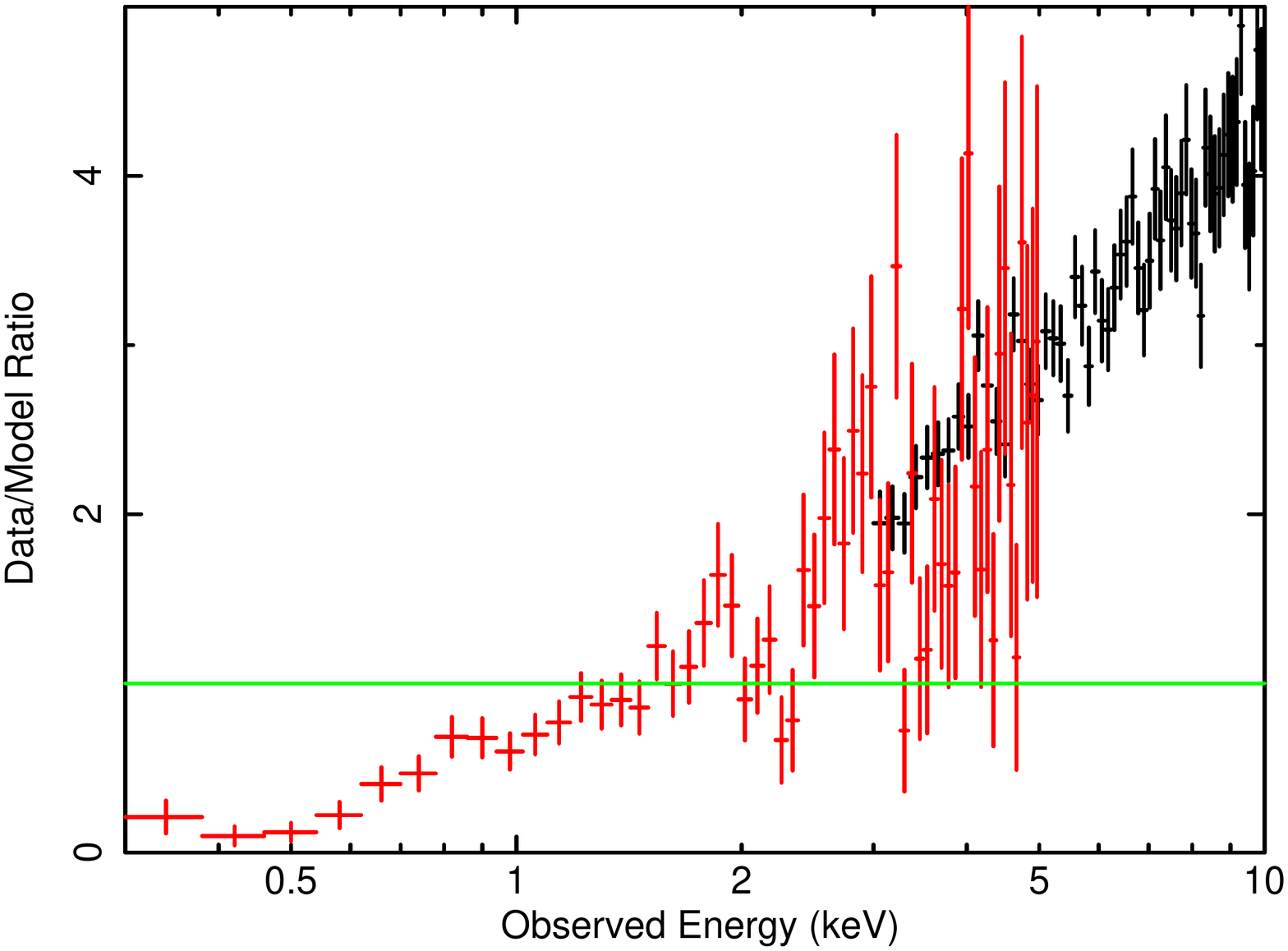}
\hspace{-0.8cm}
\includegraphics[width=6.5cm]{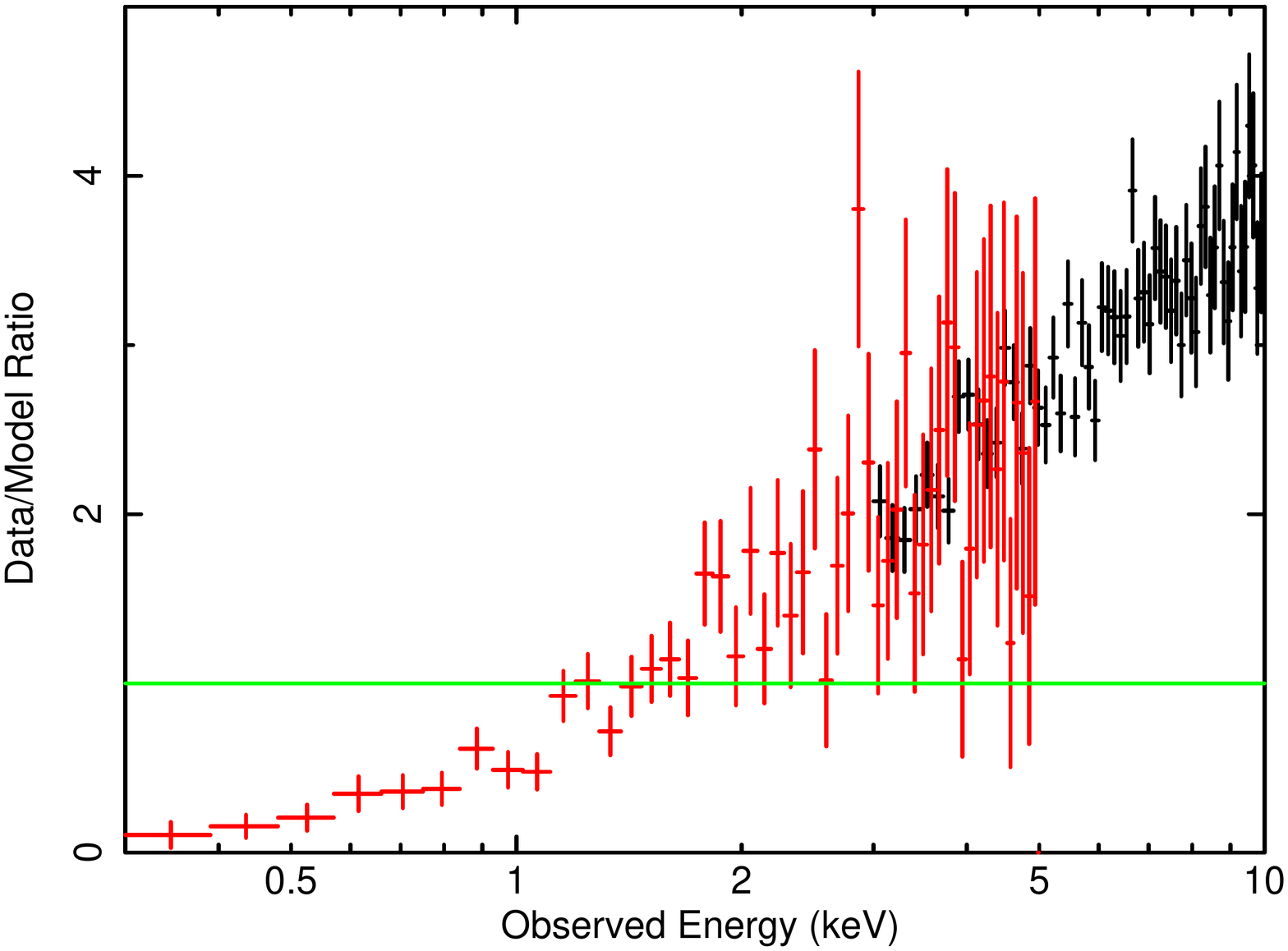}
\caption{Direct comparison of the X-ray spectra of \rbs\ at three epochs by plotting the ratio of the binned data to a $\Gamma = 2$ power law. ({\it left}) EPIC PN and MOS in 2013; ({\it middle}) \xrt\ + \nustar\ in 2014 Dec, ({\it right}) \xrt\ + \nustar\ in 2015 Jan.}
\end{figure}

\

\subsection{The RGS Spectrum} 
\label{subsec:rgs}

If the majority of the observed column density of \rbs\ is localized in one IGM filament, the spectrum is likely to show detectable absorption lines. However, if lines are not visible in the spectrum, we can conclude that such a localized absorber is not present, and the curvature is due to some other cause. For example, multiple absorbers at various redshifts would result in a collection of weak absorption lines, distributed in redshift space, which would not be detected by the X-ray gratings.
For this, we turned to the RGS, to test whether its high resolution reveals any absorption lines. The \xmm /RGS data have an average signal to noise ratio of $5.2$, $3.8$ and $0.9$ in the $6.2-12$\AA{}, $12-20$\AA{} and $20-38.2$\AA{} regions, respectively.

An attempt to obtain the redshift of the absorber by fitting for it using the tbnew model failed since the result depends heavily on the initial value of $z$.
If the absorber is at the host at $z = 2.69$, its column density ($>10^{22}\ \mathrm{cm^{-2}}$) should make for strong (redshifted) lines that would be detected. 
Figure\,5 shows the RGS spectrum as well as the prominent 1s-2p lines that would be expected from an absorber at the host redshift of $z = 2.69$, as follows: H-like S at $17.5$\AA{}, He-like S at $18.6$\AA{}, neutral S at $19.8$\AA{}, H-like Si at $22.8$\AA{}, He-like Si at $24.5$\AA{}, and neutral Si at $26.3$\AA{}. Neutral Si and S have no 1s-2p transitions, unless in an excited state with a vacancy in the 2p shell, in which case the wavelengths are as stated above. Therefore, these lines are less likely to be observed.
None of these lines is clearly discernible in the data.
Specifically, when fitting the highly ionized lines ion-by-ion \citep[using the model of][]{Peretz18}, we find that they are all consistent with zero to within $90\%$ confidence. Furthermore, we find the upper bound on the column densities of those ions to yield an equivalent $N_H$, which is significantly lower than the $N_H$ suggested by the \xmm /EPIC spectra. For example, for He-like Si we find an upper bound of $7.9\times10^{16} \mathrm{cm^{-2}}$, which is roughly equivalent to $N_H\approx10^{21} \mathrm{cm^{-2}}$, but this hydrogen would be ionized. Such highly ionized plasma would carry only a small fraction of neutral hydrogen, which is significantly lower than the $N_H=3.19\times10^{22}\mathrm{cm^{-2}}$ found based on the \xmm /EPIC data.
The neutral O line at $23.5$\AA\ resulting from the local galactic absorption is the only line detected with confidence. 
We are not able to detect any other line with high significance that also makes sense, neither with a $z = 2.69$ nor with a $z = 0$ absorber. 

A wide feature at $22.6$\AA{} can be seen in the RGS spectrum, roughly, though not exactly, at the redshifted
wavelength of H-like Si. However, if H-like Si is indeed of a high enough column density to be apparent, then so should H-like S be. The aforementioned ion-by-ion fit gives the H-like S and Si lines column densities consistent with zero and upper bounded by $5.6\times10^{16} \mathrm{cm^{-2}}$ and $1.7\times10^{18} \mathrm{cm^{-2}}$, respectively. The tighter among the two, again, gives an upper limit on $N_H$ of $\approx10^{21} \mathrm{cm^{-2}}$, which is much lower than the \xmm /EPIC result.
Furthermore, at $21.8$\AA{} a feature of similar width and depth to those of the feature at $22.6$\AA{} is clearly visible. The line at $21.8$\AA{} is, more precisely, located at $21.774 \pm{0.003}$\AA{}, which is far enough from $21.6015$\AA{}, where local, galactic O$^{+6}$ would be expected. The fact that no line is expected at $21.8$\AA{} demonstrates that effects at this scale might well be caused by the instrument or by photon count noise, rather than by real lines in the spectrum. 
Hence, we conclude that the seemingly real feature at $22.6$ \AA{} was more likely created by the instrument as well, and that no lines of a $z = 2.69$ absorber are identified in the spectrum.

The main result is that all features bar none are consistent with zero equivalent width and column density.
The absence of host absorption lines is corroborated by the lack of Ly\,$\alpha$ absorption in the optical spectrum of \rbs\ \citep{Ellison08}.
Conversely, several low-column IGM Ly\,$\alpha$ absorption systems can be seen in the figure of that paper between $z = 2.65 - 1.63$.

\begin{figure}[h]

\includegraphics[width=0.45\columnwidth]{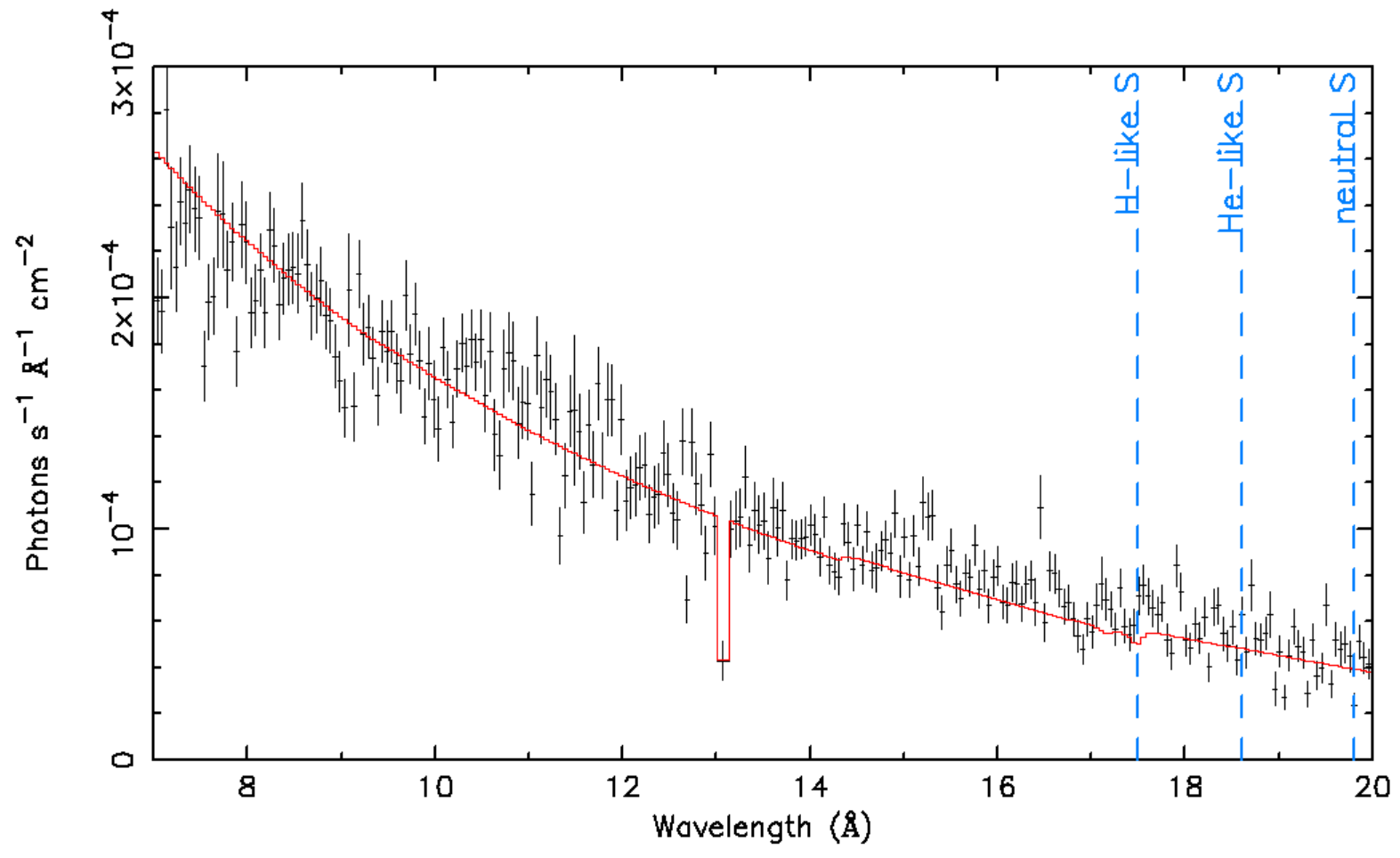}
\hspace{1cm}
\includegraphics[width=0.45\columnwidth]{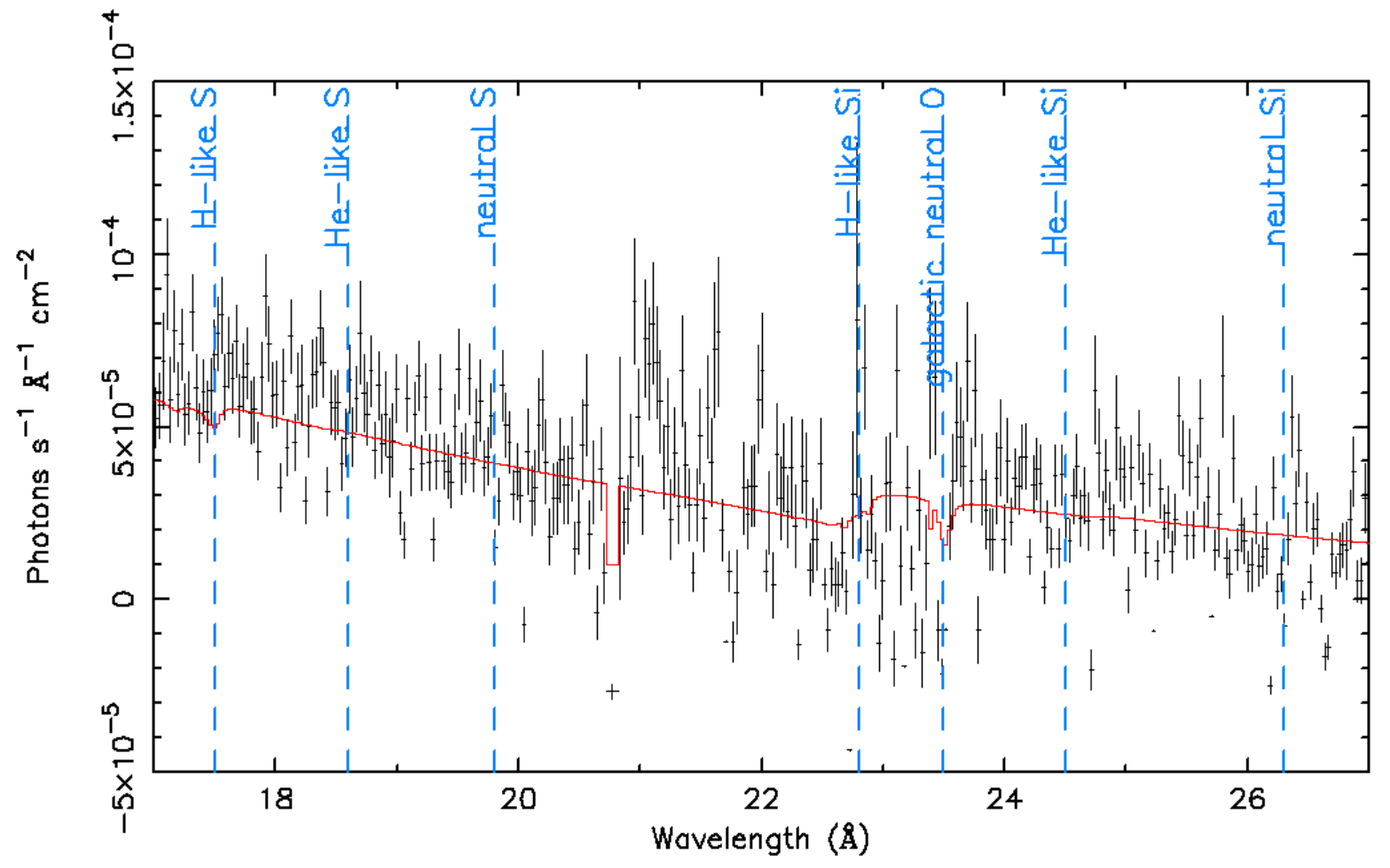}

\caption{The spectrum observed by RGS from $7$ \AA{} to $27$ \AA{}. The lines expected from an absorber located at $z=2.69$ are marked as vertical dashed lines. The solid line is the fitted model of a power law with galactic absorption and an absorber at the host at $z = 2.69$.
}

\end{figure}

\section{Conclusions} \label{sec:conclusions}

We analyzed X-ray spectra of \rbs\ from three different telescopes and at different times in an attempt to explain the spectral turnover  at low energies below 1~keV. 
Despite the extremely high S/N ratio of the \xmm /EPIC spectrum, statistical measures of model fitting to CCD spectra do not break the degeneracy between different models, which all yield similar-quality fits.
An alleged absorber is equally probable to be Galactic or extra-galactic, or non-existent if the source is intrinsically curved.
Moreover, a double broken power law with breaks at low (1\,keV) and at high ($\gtrsim5$\,keV) energy also provides a good fit.

The high resolution spectrum measured with \xmm /RGS is also of limited use, as it does not reveal any unambiguous absorption lines that could have hinted at the origin of the absorber.
The lack of X-ray spectral lines, none of which have been found to date in high-$z$ sources makes a localized absorber unlikely, regardless of the absorber redshift.
Variability remains an important tool, and although the spectrum of \rbs\ clearly varies between observations, we find that its absorber does not.
Two additional properties of the blazar population at large need to be considered.
One, local blazars do not feature similar absorbed spectra.
Two, the turn over in high-$z$ X-ray sources tends to concentrate around 1\,keV in the observed frame \citep{key-4}, regardless of the redshift of the source.
These two pieces of circumstantial evidence tend to leave intergalactic absorption as the more likely interpretation of the curvature in X-ray spectra of high-$z$ sources, which is begging for better support than could be extracted from the present spectra of \rbs .



\end{document}